\documentclass[a4paper,fleqn,usenatbib]{mnras}

\usepackage[T1]{fontenc}
\usepackage{ae,aecompl}

\usepackage{graphicx}	
\usepackage{amsmath}	
\usepackage{amssymb}	

\title[Unbeamed properties of blazar 4C~71.07]{Unveiling the monster heart: unbeamed properties of blazar 4C~71.07}
\author[C. M. Raiteri et al.]{
C.~M.~Raiteri$^{1}$\thanks{E-mail: claudia.raiteri@inaf.it},
J.~A.~Acosta Pulido$^{2,3}$,
M.~Villata$^{1}$,
M.~I.~Carnerero$^{1}$,
\newauthor
P.~Romano$^{4}$,
and S.~ Vercellone$^{4}$
\\
$^{1}$INAF, Osservatorio Astrofisico di Torino, via Osservatorio 20, I-10025 Pino Torinese, Italy\\
$^{2}$Instituto de Astrofisica de Canarias (IAC), La Laguna, E-38200 Tenerife, Spain\\
$^{3}$Departamento de Astrofisica, Universidad de La Laguna, La Laguna, E-38205 Tenerife, Spain\\
$^{4}$INAF, Osservatorio Astronomico di Brera, Via Emilio Bianchi 46, 23807 Merate, LC, Italy
}

\date{Accepted XXX. Received YYY; in original form ZZZ}

\pubyear{2018}

\begin{document}
\label{firstpage}
\pagerange{\pageref{firstpage}--\pageref{lastpage}}
\maketitle

\begin{abstract}
4C~71.07 is a high-redshift blazar whose optical radiation is dominated by quasar-like nuclear emission. We here present the results of a spectroscopic monitoring of the source to study its unbeamed properties. We obtained 24 optical spectra at the Nordic Optical Telescope (NOT) and William Herschel Telescope (WHT) and 3 near-infrared spectra at the Telescopio Nazionale Galileo (TNG). They show no evidence of narrow emission lines. The estimate of the systemic redshift from the \ion{H}{$\beta$} and \ion{H}{$\alpha$} broad emission lines leads to $z_{\rm sys}=2.2130 \pm 0.0004$. Notwithstanding the nearly face-on orientation of the accretion disc, the high-ionization emission lines present large broadening as well as noticeable blueshifts, which increase with the ionizing energy of the corresponding species. This is a clear indication of strong ionized outflows. Line broadening and blueshift appear correlated. We applied scaling relationships to estimate the mass of the supermassive black hole from the Balmer and \ion{C}{IV} lines, taking into account the prescriptions to correct for outflow. They give $M_{\rm BH} \sim 2 \times 10^9 \, M_\odot$. We derived an Eddington luminosity $L_{\rm Edd} \sim 2.5 \times 10^{47} \rm \, erg \, s^{-1}$ $\sim L_{\rm disc}$, and a broad line region luminosity of $L_{\rm BLR} \sim 1.5 \times 10^{46} \rm \, erg \, s^{-1}$.
The line fluxes do not show significant variability in time. In particular, there is no line reaction to the jet flaring activity detected in 2015 October--November. This implies that the jet gives no contribution to the photoionization of the broad line region in the considered period.
\end{abstract}




\section{Introduction}
Active galactic nuclei (AGN) come in many flavours, but all of them are thought to be powered by accretion of matter onto a supermassive black hole. The emission from the accretion disc is observed as a ``big blue bump" in the source spectral energy distribution (SED) and peaks in the rest-frame ultraviolet band. Fast moving gas clouds outside the disc form the broad line region (BLR), producing broad emission lines with velocities of several thousand $\rm km \, s^{-1}$. On larger spatial scales, slower gas clouds build up the narrow line region (NLR), emitting lines with velocities of the order of several hundred $\rm km \, s^{-1}$. Radio-loud AGN are also characterized by two plasma jets ejected from the poles of the central black hole. If one of these jets is oriented close to the line of sight, we name the object "blazar". 
Therefore, blazars are a relatively rare population of radio-loud AGN, numbering a few thousand known objects. The emission from the relativistic jet pointing toward us undergoes Doppler boosting and beaming, so that blazar flux is enhanced and the variability time scales are contracted \citep[see e.g.][]{urry1995}. Because of this, the non-thermal emission from the jet (radiation due to synchrotron process at low energies, and to inverse-Compton and/or hadronic processes at high energies) usually dominates on the other emission contributions, coming from the host galaxy or the nuclear region (accretion disc, BLR and NLR). The jet emission can be so strong that it completely overwhelms the nuclear component. 
When the nuclear emission is weak or even undetected we classify the blazar as a BL Lac object, while it is named flat-spectrum radio quasar (FSRQ) when the ``quasar core" can be observed \citep{stickel1991,stocke1991,giommi2012}.
The distinction between BL Lac objects and FSRQ is however a still debated issue \citep[e.g.][]{ghisellini2011}.

Blazars are mostly studied because they allow us to shed light on the emission properties of the extragalactic jets over the whole electromagnetic spectrum, from the radio band up to the most energetic $\gamma$ rays at TeV energies. Since their jets are formidable particle accelerators, they are also candidate sources of high-energy cosmic rays and neutrinos \citep{ice2018a,ice2018b}.

Though less popular, the study of the nuclear, unbeamed properties of blazars is also interesting, because it can tackle a number of key issues.
For example, their relationship with other classes of AGN, in particular unbeamed quasars.
The parent population of blazars has been identified with radio galaxies. FR I and FR II would likely be the counterpart of BL Lacs and FSRQ, respectively, but this correspondence is questioned by several observational inconsistencies \citep[e.g.][]{urry1995}.

It is also not clear yet if, in addition to the thermal radiation from the disc, also the jet  photons play a role in photoionizing the BLR. A number of works attempted to detect line flux variability correlated with continuum flux changes, where the continuum is mainly jet emission, and obtained different results \citep{corbett2000, raiteri2007a,leon2013,isler2013,carnerero2015,isler2015}. It seems that the BLR can react to some jet flares, but this is not a common behaviour.

Moreover, as the orientation of blazars is known, the estimate of gas velocities from the spectral lines can offer a good test for the models predicting BLR and NLR geometries. The same is true for the understanding of AGN feedback on the host galaxy evolution. Many AGN show evidence of outflow on different spatial scales. 
While narrow emission lines, chiefly [\ion{O}{III}], can shed light on gas dynamics over kpc scales, broad emission lines can probe the inner nuclear zone, where the outflows are launched from.
The presence of outflows in quasar spectra is revealed by asymmetries in the line profiles and/or shifts of the emission lines with respect to the systemic redshift, which represents the redshift of an ideal light source at rest in the galaxy gravity centre. This is usually measured from low-excitation broad lines (usually \ion{H}{$\beta$} in low-redshift objects and \ion{Mg}{II} in high-$z$ ones), or from narrow emission lines, mainly [\ion{O}{III}] $\lambda 5007$. Strong outflows can make the determination of the systemic redshift problematic \citep[e.g][]{perrotta2019} because even if blueshifts are larger in broad emission lines of high-excitation species, they can also affect the other lines, in particular [\ion{O}{III}].   

\citet{vandenberk2001} analysed more than 2200 quasar spectra from the SDSS. They found that narrow forbidden lines have shifts smaller than $\rm 100 \, km \, s^{-1}$, while permitted and semiforbidden lines can reach $550 \rm \, km \, s^{-1}$. They confirmed earlier suggestions that the absolute value of the velocity offset of the emission lines relative to the laboratory rest wavelengths grows with the ionization potential. 
In contrast, \citet{meyer2019} found that while the high-ionization \ion{C}{IV} line is strongly blueshifted with respect to the low-ionization lines \ion{O}{I}, \ion{C}{II} and \ion{Mg}{II}, the intermediate-ionization \ion{Si}{IV} and \ion{C}{III]} are not. They also found that the \ion{C}{IV} blueshift shows no significant evolution up to $z \sim 6$, while it abruptly increases at higher redshifts.

The FSRQ 4C~71.07 (0836+710) is an ideal target to study the nuclear emission of blazars. Its SED shows a prominent big blue bump \citep{raiteri2014, raiteri2019} and its spectra reveal strong emission lines. A redshift $z=2.172$ was estimated by \citet{stickel1993} from the \ion{C}{IV} $\lambda1549$ and \ion{C}{III]} $\lambda1909$ broad emission lines. These authors also identified an intervening absorption system at $z=0.914$ from an absorption line at $\sim 5360$ \AA, which was attributed to \ion{Mg}{II} $\lambda\lambda2796,2803$. This was later confirmed by \citet{scott2000}, who also found a number of other absorbers at redshifts $z=1.4256$, 1.6681, 1.7331, and 2.1800. 
A value of $z=2.18032$ for the source redshift was obtained by \citet{lawrence1996} from spectroscopic observations made with the Palomar telescope in 1983.
By fitting the [\ion{O}{III}] $\lambda 5007$ narrow line in $H$-band spectra, \citet{mcintosh1999} estimated a systemic redshift of $z=2.218$. 
The difference in the redshift estimates obtained with different lines is by itself a symptom of possible outflows that deserves further detailed analysis.

In 2014--2016 the Whole Earth Blazar Telescope\footnote{http://www.oato.inaf.it/blazars/webt/} \citep[WEBT, e.g.][]{villata2002,villata2006,raiteri2017_nature} organized a multiwavelength observing campaign to study both the beamed and unbeamed properties of 4C~71.07. The results of the optical, near-infrared, and radio observations by the WEBT members, complemented by ultraviolet and X-ray data from the {\it Swift} satellite and $\gamma$-ray data from the {\it Fermi} spacecraft, are presented in \citet{raiteri2019}. 
During the whole campaign, we performed an optical spectroscopic monitoring 
and obtained some near-infrared spectroscopic observations. 
We here analyse in detail these data to unveil the properties of the 4C~71.07 quasar core.

The dataset is presented in Sect.\ \ref{observations}. In Sect.\ \ref{redshift} we estimate the systemic redshift from the \ion{H}{$\beta$} and \ion{H}{$\alpha$} emission lines. The data analysis procedure is described in Sect.\ \ref{analysis} and applied to the Balmer lines in Sect.\ \ref{balmer}, and to the high-excitation lines in Sect.\ \ref{hil}. Section \ref{ratios} compares our line flux ratios with those characterising various quasar spectra templates. 
In Sect.\ \ref{bh} we applied scaling relationships to derive the black hole mass.
The BLR radius and luminosity are calculated in Sect.\ \ref{blr} and compared with the disc and Eddington luminosities. Section \ref{blue} deals with the line blueshifts and outflow properties.  Line flux variability is discussed in Sect.\ \ref{lineva}. Finally, we outline our conclusions in Sect.\ \ref{fine}.

In this paper we assume a flat Universe with $H_0= 70 \rm \, km \, s^{-1} Mpc^{-1}$, $\Omega_m=0.3$, $\Lambda_0=0.7$, $q_0=-0.55$.

\section{Spectroscopic observations}
\label{observations}

During the multifrequency WEBT campaign of 2014--2016 \citep{raiteri2019}, we organized a spectroscopic monitoring with the intent of investigating the unbeamed properties of 4C~71.07. 
Optical spectra were acquired with the 4.2 m William Herschel Telescope (WHT) and 2.6 m Nordic Optical Telescope (NOT), both located at the Roque de los Muchachos Observatory in the Canary Islands. A list of the observations is given in Table \ref{spettrilog}.
We chose instrumental configurations with a large wavelength coverage to include the strongest broad emission lines.

\begin{table}
	\centering
	\caption{Log of the spectroscopic observations of 4C 71.07.}
	\label{spettrilog}
	\begin{tabular}{llcc} 
		\hline
		Date & JD$^a$ & Telescope & Instrument\\
		\hline
                2014 12 05 & 6997.751 & WHT & ACAM   \\ 
                2014 12 27 & 7019.499 & WHT & ACAM   \\
                2015 01 29 & 7052.429 & WHT & ACAM   \\ 
		2015 02 12 & 7066.389 & NOT & ALFOSC \\ 
                2015 03 03 & 7085.415 & WHT & ACAM   \\ 
                2015 03 04 & 7086.522 & WHT & ACAM   \\
                2015 04 20 & 7133.453 & WHT & ACAM   \\ 
		2015 04 24 & 7137.463 & NOT & ALFOSC \\
                2015 06 02 & 7176.400 & WHT & ACAM   \\ 
		2015 06 17 & 7191.411 & NOT & ALFOSC \\ 
		2015 07 16 & 7220.393 & NOT & ALFOSC \\ 
		2015 08 21 & 7255.640 & NOT & ALFOSC \\ 
                2015 10 06 & 7302.734 & WHT & ACAM   \\ 
                2015 11 13 & 7340.711 & NOT & ALFOSC \\ 
                2016 01 07 & 7395.443 & WHT & ACAM   \\ 
                2016 01 18 & 7406.622 & WHT & ACAM   \\
                2016 04 01 & 7480.485 & TNG & NICS   \\ 
                2016 04 06 & 7485.426 & NOT & ALFOSC \\
                2016 05 17 & 7526.435 & NOT & ALFOSC \\ 
                2016 06 01 & 7541.486 & NOT & ALFOSC \\ 
                2016 07 03 & 7573.492 & NOT & ALFOSC \\ 
                2016 08 04 & 7605.648 & NOT & ALFOSC \\ 
                2016 09 04 & 7636.730 & NOT & ALFOSC \\ 
                2016 10 11 & 7672.702 & WHT & ACAM   \\ 
                2016 10 20 & 7682.707 & WHT & ACAM   \\
                2016 10 28 & 7690.697 & TNG & NICS   \\
                2019 04 11 & 8586.450 & TNG & NICS   \\
		\hline
	        $^a \, \rm Julian \, Date - 2450000$
\end{tabular}
\end{table}

Spectroscopic observations at the NOT were performed with the ALFOSC camera and grism 4, ranging from 3200 to 9100 \AA, and resolution $R \sim 360$ for a 1 arcsec slit width.
We adopted the atmospheric dispersion corrector (ADC) to compensate for flux losses, which were significant in July--August, when the source was observed at high air masses.
The WHT spectra were taken with the ACAM instrument, covering the wavelength range 3500--9400 \AA, and resolution $R \sim 450$ for a 1 arcsec slit width. 
The source spectra were calibrated with respect to the spectroscopic standard star BD+75 325. 

To check and improve flux calibration, we used source photometry obtained either from the same telescope just before/after the spectroscopic observations, or from the other telescopes participating to the WEBT campaign, within the observing night. 

\begin{figure}
	\includegraphics[width=\columnwidth]{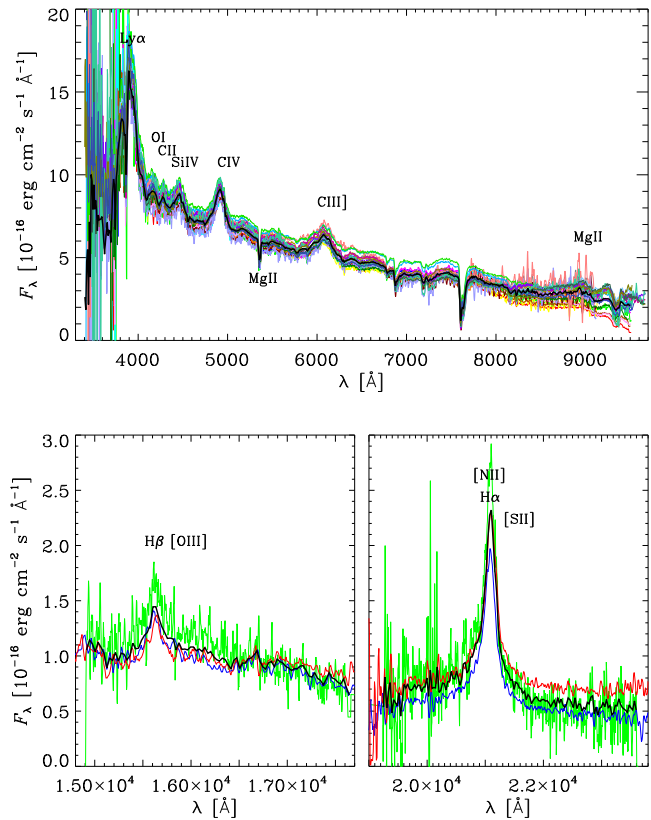}
    \caption{Top: Optical spectra of 4C~71.07 taken with the WHT and NOT telescopes.
The broad emission lines Ly$\alpha$, \ion{O}{I}, \ion{C}{II}, \ion{Si}{IV}, \ion{C}{IV}, and \ion{C}{III]} are clearly visible, together with the \ion{Mg}{II} absorption line at $z=0.914$. The \ion{Mg}{II} broad emission line is corrupted by telluric absorption.
Bottom: Near-infrared spectra acquired at the TNG telescope, showing the \ion{H}{$\beta$} (left) and \ion{H}{$\alpha$} (right) Balmer lines.
Thick black lines represent mean spectra.}
    \label{medio}
\end{figure}

The observed calibrated spectra are shown in Fig.\ \ref{medio}, where the broad emission lines \ion{Ly}{$\alpha$}, \ion{O}{I}, \ion{C}{II}, \ion{Si}{IV}, \ion{C}{IV}, and \ion{C}{III]} are clearly visible, together with the \ion{Mg}{II} absorption line at $z=0.914$ identified by \citet{stickel1993}. The \ion{Mg}{II} emission line is strongly affected by telluric absorption by $\rm H_2O$ molecules \citep{rudolf2016}.
The \ion{Ly}{$\alpha$} profile is almost split in two by the onset of the \ion{Ly}{$\alpha$} forest, the series of blueward absorption lines mainly due to intervening neutral hydrogen, which is usually observed in high-redshift quasars \citep{scott2000}.

During the WEBT campaign we also acquired two near-infrared spectra with the 3.6 m Telescopio Nazionale Galileo (TNG) at the Roque de los Muchachos Observatory. The near-infrared observations were performed with the NICS instrument and the HK disperser. 
A further spectrum was obtained with the same telescope and detector well after the end of the WEBT campaign to improve the data analysis (see Table \ref{spettrilog}). 
The observed near-infrared calibrated spectra are shown in Fig.\ \ref{medio}. As in the optical case, calibration was checked against photometry. The Balmer \ion{H}{$\beta$} and \ion{H}{$\alpha$} lines are well defined, but the narrow [\ion{O}{III}] lines redward of \ion{H}{$\beta$}, the narrow [\ion{N}{II}] lines blended with \ion{H}{$\alpha$}, and the narrow [\ion{S}{II}] redward of \ion{H}{$\alpha$} are not clearly distinguishable.

Data reduction was performed using Image Reduction and Analysis Facility (IRAF) standard routines for spectroscopic data. All spectra were subsequently analysed with programs written in Interactive Data Language (IDL). They were first corrected for Galactic extinction according to \citet{fitzpatrick1999}, with $E(B-V)=0.026$ and $R_V=3.1$.

\section{Systemic redshift determination}
\label{redshift}

As mentioned in the Introduction, an estimate of the systemic redshift of 4C~71.07 was obtained by \citet{mcintosh1999} from $H$-band spectra acquired in 1993 with the 4.5 m Multiple Mirror Telescope. They derived $z=2.218$ from modelling a possible [\ion{O}{III}] line, assuming a very smooth and broad Fe contribution. They also stressed the difference in $z$ values obtained from [\ion{O}{III}] and \ion{H}{$\beta$}.
An extremely weak contribution of the NLR is a common feature in high-redshift quasars \citep[e.g.][]{netzer2004,vietri2018}. Moreover, the equivalent width (EW) of [\ion{O}{III}] has been proposed as an orientation indicator, with more face-on objects presenting weaker [\ion{O}{III}] lines \citep{risaliti2011,bisogni2017}. 4C~71.07 is a high-redshift blazar, so it is not surprising that in our near-infrared spectra, the [\ion{O}{III}] line is not clearly recognisable. Instead, the Balmer lines are well defined.
Therefore, we consider a systemic redshift estimate based on Balmer lines as more secure for this source. The absence of visible narrow lines likely implies that the [\ion{N}{II}] blending on the \ion{H}{$\alpha$} is a second-order effect and that this line can also be used in addition to \ion{H}{$\beta$} for a systemic redshift determination.

\begin{figure}
	\includegraphics[width=\columnwidth]{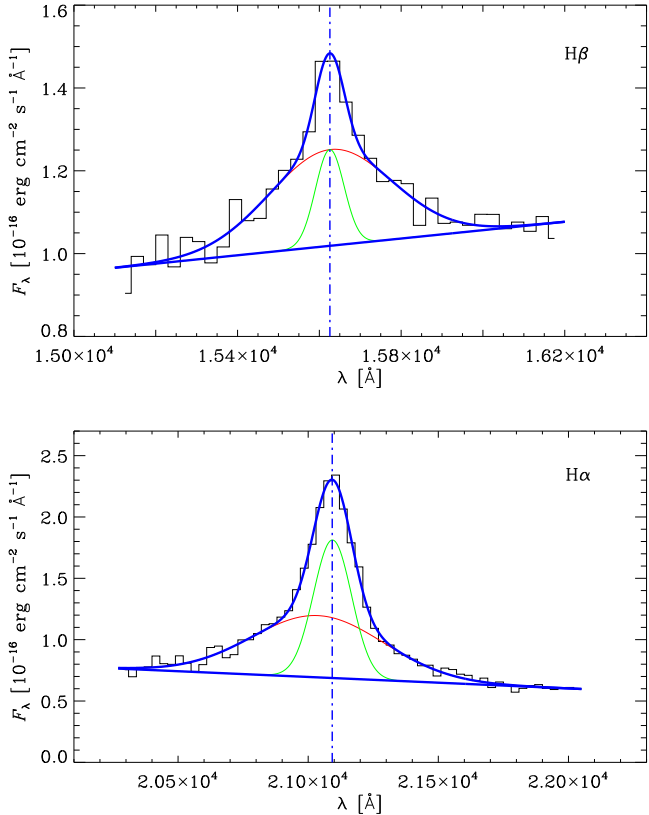}
    \caption{Fits to the \ion{H}{$\beta$} (top) and \ion{H}{$\alpha$} (bottom) line profiles of the mean near-infrared dereddened spectrum. The model includes 2 Gaussians and a linear continuum. The vertical lines mark the peak of the continuum-subtracted model, from which a source redshift of $z=2.2130 \pm 0.0004$ is inferred.}
    \label{redsys}
\end{figure}

To improve the S/N, we consider the mean near-infrared spectra shown in Fig.\ \ref{medio}. After reddening correction, we fit them with multiple Gaussians on a linear continuum. We find that a combination of two Gaussians leads to a good fit of the line profiles of both \ion{H}{$\beta$} and \ion{H}{$\alpha$} (see Fig.\ \ref{redsys}). 
We estimate the redshift from the peaks of the continuum-subtracted models. The uncertainty is evaluated through 1000 Monte Carlo simulations, where the flux density of each point in the spectrum is perturbed randomly according to a Gaussian distribution of the error. This is taken as the standard deviation of the flux densities in a spectral region close to the line.
The results are $z=2.2134 \pm 0.0007$ and $z=2.2129 \pm 0.0003$ from the \ion{H}{$\beta$} and \ion{H}{$\alpha$} lines, respectively.
A weighted mean finally gives $z_{\rm sys}=2.2130 \pm 0.0004$, which we assume as the source systemic redshift.

\section{Data analysis}
\label{analysis}

We used our systemic redshift determination, $z_{\rm sys}=2.213$, to shift the spectra to the rest-frame.
The flux density was transformed accordingly: $$F_{\lambda_{\rm rest}}=F_{\lambda_{\rm obs}} \times (\lambda_{\rm obs} / \lambda_{\rm rest})=(1+z_{\rm sys}) \, F_{\lambda_{\rm obs}},$$
where $\lambda_{\rm rest}$ and $\lambda_{\rm obs}$ are the rest-frame and observed wavelengths, respectively.

Because of the known anti-correlation between the \ion{Fe}{II} and [\ion{O}{III}] in quasar spectra \citep[e.g.][]{boroson1992}, we expect an important iron contribution in this object. 
We considered the Fe templates by \cite{vestergaard2001} for the optical (rest-frame UV) spectra, and by \citet{veron2004} for the near-infrared (rest-frame optical) ones. 
The templates were conveniently scaled and broadened to match the spectral behaviour. 
We had to apply different scaling factors in different wavelength windows, as expected \cite[e.g.][]{boroson1992,vestergaard2001,tsuzuki2006}. 
In general, the main features of the \ion{Fe}{II} templates are matched in wavelength, implying that \ion{Fe}{II} emission is not affected by outflows/inflows or, at least, it is subject to the same amount of blueshift/redshift as the \ion{H}{$\beta$} and \ion{H}{$\alpha$} lines. This in turn implies that \ion{Fe}{II} emission likely comes from the outer zones of the BLR, where the Balmer lines are thought to be produced. In contrast, we found a blueshift of the high-ionization \ion{Fe}{III} line (see Sect.\ \ref{hil}).

We identified small spectral regions as line free as possible to derive power-law fits $F_\lambda \propto \lambda^{\alpha_\lambda}$ to the continuum. This was done separately in the optical and in the two near-infrared regions around the \ion{H}{$\beta$} and \ion{H}{$\alpha$} lines. 
These small spectral regions were also used to estimate the mean uncertainty on the flux densities.
We subtracted the Fe and continuum contributions from the spectra and performed multiple Gaussian fitting on the residuals.

We already mentioned that the estimate of different redshift values from different lines in the literature suggests the presence of blueshifts due to outflows. Therefore, we introduce the blueshift velocity, which is defined as:
\begin{equation}
v_{\rm BS}=c \, {{\lambda^{\rm peak}_{\rm obs} - \lambda_{\rm sys}} \over {\lambda_{\rm sys}}} = 
c \, {{\lambda^{\rm peak}_{\rm rest} - \lambda_{\rm lab}} \over {\lambda_{\rm lab}}} = 
c \, {{z_{\rm obs}-z_{\rm sys}} \over {1+z_{\rm sys}}}
\label{bs}
\end{equation}
where $c$ is the speed of light, $z_{\rm obs}$ is the redshift estimate according to the line model peak $\lambda^{\rm peak}_{\rm obs}=(1+z) \, \lambda^{\rm peak}_{\rm rest}$, and $\lambda_{\rm sys}=(1+z) \, \lambda_{\rm lab} $ the central wavelength that the line would have at the systemic redshift $z_{\rm sys}$.

In the following, we will discuss the spectral fitting of the main emission lines in the mean near-infrared and optical spectra. The main results are reported in Table \ref{results}, where uncertainties on the estimated parameters were calculated with the same Monte Carlo flux-redistribution method adopted in Section \ref{redshift} to determine the error on the systemic redshift.
In table \ref{results}, Column 1 lists the line, Column 2 its laboratory wavelength $\lambda_{\rm lab}$, Column 3 the peak of the line profile model in the rest frame $\lambda^{\rm peak}_{\rm rest}$ (according to the systemic redshift $z_{\rm sys}=2.213$), Column 4 the redshift derived from the line model peak $z_{\rm obs}$, Column 5 the line flux obtained by integrating the line profile model, Column 6 the isotropic luminosity assuming a luminosity distance of 17585 Mpc, Column 7 the velocity estimated from the FWHM of the line model after correction for instrumental broadening ($\rm FWHM=\sqrt{FWHM^2_{obs}-FWHM^2_{inst}}$), and Column 8 the blueshift velocity defined in Eq.\ \ref{bs}. We note that the velocity of the BLR clouds is $v_{\rm BLR}= f \times v_{\rm FWHM}$, where $f$ is an unknown factor depending on the structure and geometry of the BLR. 
Different values of $f$ can be obtained when analysing different lines. Observations of the \ion{Mg}{II} line suggest that the low-ionization BLR has a disc-like geometry, while the \ion{C}{IV} line would indicate a more isotropic high-ionization BLR \citep{fine2011}.

\begin{table*}
	\centering
	\caption{Results of the spectral fitting of the main broad emission lines in our mean spectra.}
	\label{results}
	\begin{tabular}{lccccccc} 
		\hline
Line   & $\lambda_{\rm lab}$ & $\lambda^{\rm peak}_{\rm rest}$ & $z_{\rm obs}$ & $F$ & $L$ & $v_{\rm FWHM}$ & $v_{\rm BS}$ \\
       & [\AA]               & [\AA]               &     & [$10^{-15} \rm \, erg \, cm^{-2} \, s^{-1}$]  & [$10^{44}\rm \, erg \, s^{-1}$] & [$\rm km \, s^{-1}$] & [$\rm km \, s^{-1}$]\\
\hline
\ion{O}{I}        & 1304.35 & $1303.06 \pm 1.45$ & $2.2098 \pm 0.0036$ & $ 3.89 \pm 0.50$ & 1.43 & $3510 \pm 362$ & $-296 \pm 334$\\
\ion{C}{II}       & 1335.30 & $1331.70 \pm 0.70$ & $2.2043 \pm 0.0017$ & $ 3.90 \pm 1.03$ & 1.44 & $3408 \pm 743$ & $-808 \pm 157$\\
\ion{Si}{IV}      & 1396.76 & $1390.20 \pm 0.96$ & $2.1979 \pm 0.0022$ & $13.35 \pm 0.99$ & 4.92 & $6827 \pm 352$ & $-1408 \pm 205$\\
\ion{C}{IV}       & 1549.06 & $1530.41 \pm 0.28$ & $2.1743 \pm 0.0006$ & $40.66 \pm 0.56$ & 15.0 &$7451 \pm 205$ & $-3609 \pm 54$\\ 
\ion{C}{III]}     & 1908.73 & $1891.75 \pm 1.71$ & $2.1844 \pm 0.0029$ & $22.57 \pm 2.16$ & 8.32 &$8173 \pm 552$ & $-2667 \pm 268$\\ 
\ion{H}{$\beta$}  & 4862.68 & $4862.99 \pm 0.23$ & $2.2132 \pm 0.0002$ & $13.71 \pm 0.78$ & 5.05 &$3240 \pm 572$ & $19 \pm 14$\\  
\ion{H}{$\alpha$} & 6564.61 & $6564.64 \pm 0.04$ & $2.21301 \pm 0.00001$ & $57.81 \pm 1.41$ & 21.3 &$3131 \pm 119$ & $1.4  \pm 0.1$\\  
		\hline
        \end{tabular}
\end{table*}

\section{Balmer lines}
\label{balmer}

\subsection{\ion{H}{$\beta$}}
\begin{figure}
	\includegraphics[width=\columnwidth]{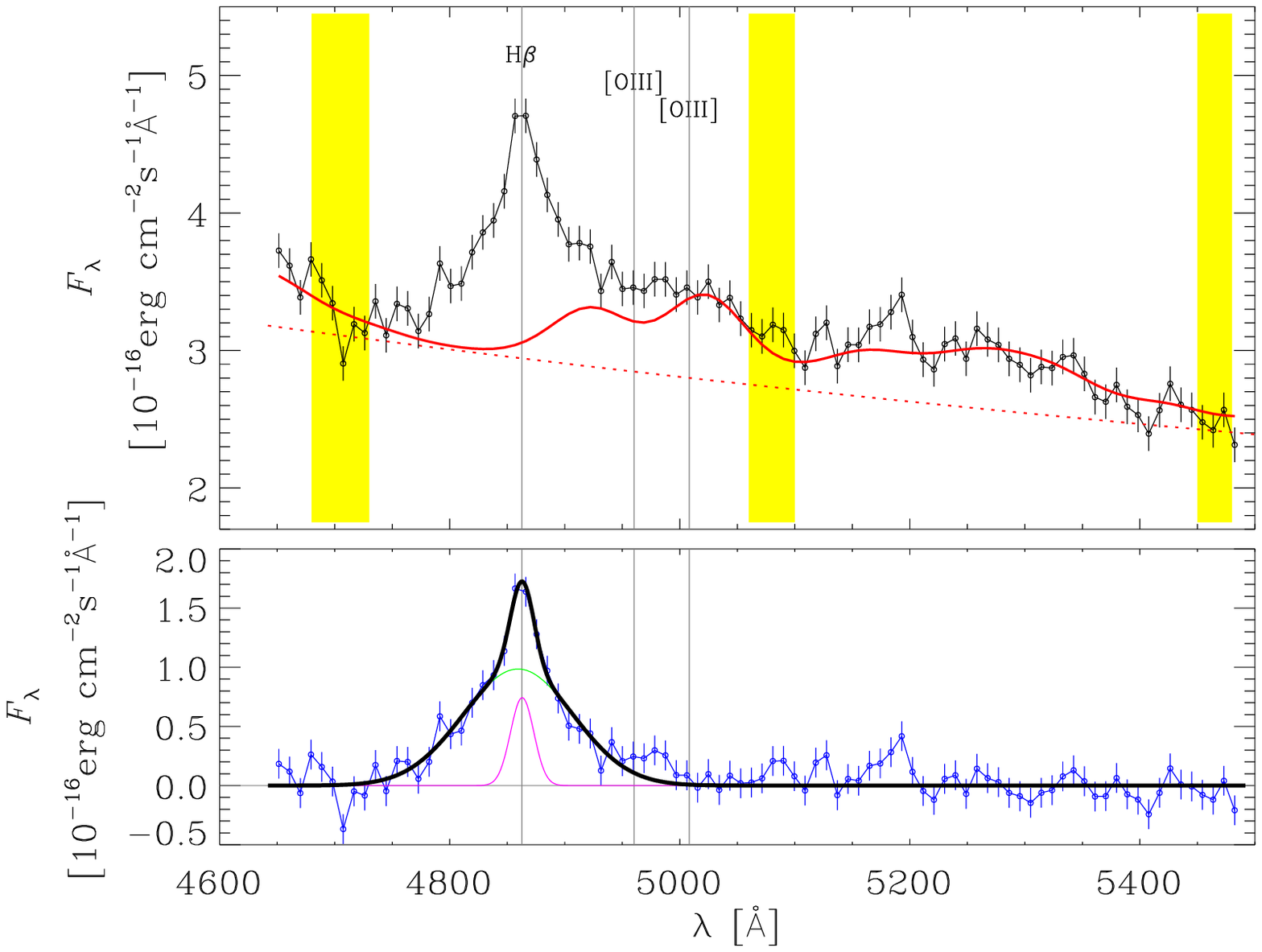}
    \caption{The \ion{H}{$\beta$} line spectral region of the mean near-infrared dereddened spectrum, shifted to the rest frame according with $z=2.213$. Top: the black points show the spectrum before the continuum and Fe subtraction. The continuum (red dotted line) is obtained by fitting the spectral zones marked with yellow stripes with a power law.
The zones in yellow were also used to estimate the mean flux uncertainty. 
The red solid line highlights the Fe contribution, which is obtained by scaling and broadening the \citet{veron2004} Fe template. Bottom: the blue points show the clean spectrum, after removal of Fe and continuum. Magenta and green lines are the two Gaussians that give the best-fit model of the line, shown in black. Vertical grey lines indicate the position of the \ion{H}{$\beta$} line and [\ion{O}{III}] doublet at $\lambda_{\rm lab}$.}
    \label{fitto_hb}
\end{figure}

Figure \ref{fitto_hb} shows the results of our spectral fitting to the \ion{H}{$\beta$} line in the mean near-infrared spectrum.
The slope of the continuum is $\alpha_\lambda=-1.57$ ($\alpha_\nu=-0.43$), which is very close to the value found by \citet{vandenberk2001} for wavelengths up to this line.
The Fe- and continuum-subtracted spectrum is fitted with two Gaussians.
As mentioned in Sect.\ \ref{redshift}, there is no hint for the presence of the [\ion{O}{III}] $\lambda 5007$ line, so we assume that the contribution of the NLR is negligible. According to the results of \citet{bisogni2017}, a FWHM velocity of about $3200 \rm \, km \, s^{-1}$ of the  
\ion{H}{$\beta$} line in composite quasar spectra implies a very low EW of the [\ion{O}{III}] line of the order of 3 \AA. The \citet{bisogni2017} results were found by analysing 12300 quasars from the SDSS, with redshift up to 0.8. We expect that in high-redshift objects the EW may be even smaller, further confirming our assumption.

\subsection{\ion{H}{$\alpha$}}
\begin{figure}
	\includegraphics[width=\columnwidth]{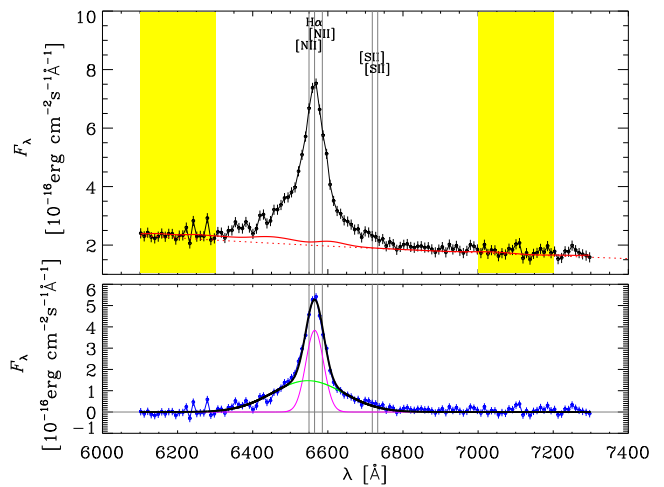}
    \caption{As in Fig.\ \ref{fitto_hb}, but for the \ion{H}{$\alpha$} line.}
    \label{fitto_ha}
\end{figure}

If the weight of the NLR is negligible, we do not have to worry about the contributions of the [\ion{N}{II}] and [\ion{S}{II}] doublets to the \ion{H}{$\alpha$} line and line red wing, respectively. 
In this region the slope of the continuum is $\alpha_\lambda=-2.03$ ($\alpha_\nu=0.03$), much harder than in the composite quasar spectrum of \citet{vandenberk2001}, who give $\alpha_\lambda=0.45$ ($\alpha_\nu=-2.45$).
This is not surprising, because the SED of those sources can receive additional softer contributions by hot-dust emission and host-galaxy light \citep{vandenberk2001}.
A good Gaussian fitting of the \ion{H}{$\alpha$} line is obtained with two components (Fig.\ \ref{fitto_ha}).

\subsection{Comparison between \ion{H}{$\beta$} and \ion{H}{$\alpha$}}

Notwithstanding the non exceptional quality of the mean spectrum in the \ion{H}{$\beta$} spectral region and the uncertainty in modelling the Fe emission in the red wing of the line, we obtained very reasonable results.

First of all, we note that spectral fitting of the Balmer lines after Fe and continuum subtraction confirms the redshift estimate obtained in Section \ref{redshift} and that, consequently, their velocity shift is essentially compatible with zero (see Table \ref{results}).

The flux ratio between \ion{H}{$\alpha$} and \ion{H}{$\beta$} is 4.2, slightly higher than the value 3.6 found in average QSO spectra \citep[e.g.][]{vandenberk2001}, but smaller than the value of about 7 found by \citet{cristiani1990}. Line flux ratios will be further discussed in Sect.\ \ref{ratios}.

In quasars, the FWHM of the broad \ion{H}{$\beta$} line is systematically larger than that of the \ion{H}{$\alpha$} line \citep{shen2011}. The values we measured in 4C~71.07 are consistent with this (see Table \ref{results}), even if our FWHMs are affected by large uncertainties.

\section{High-ionization lines}
\label{hil}

\begin{figure}
	\includegraphics[width=\columnwidth]{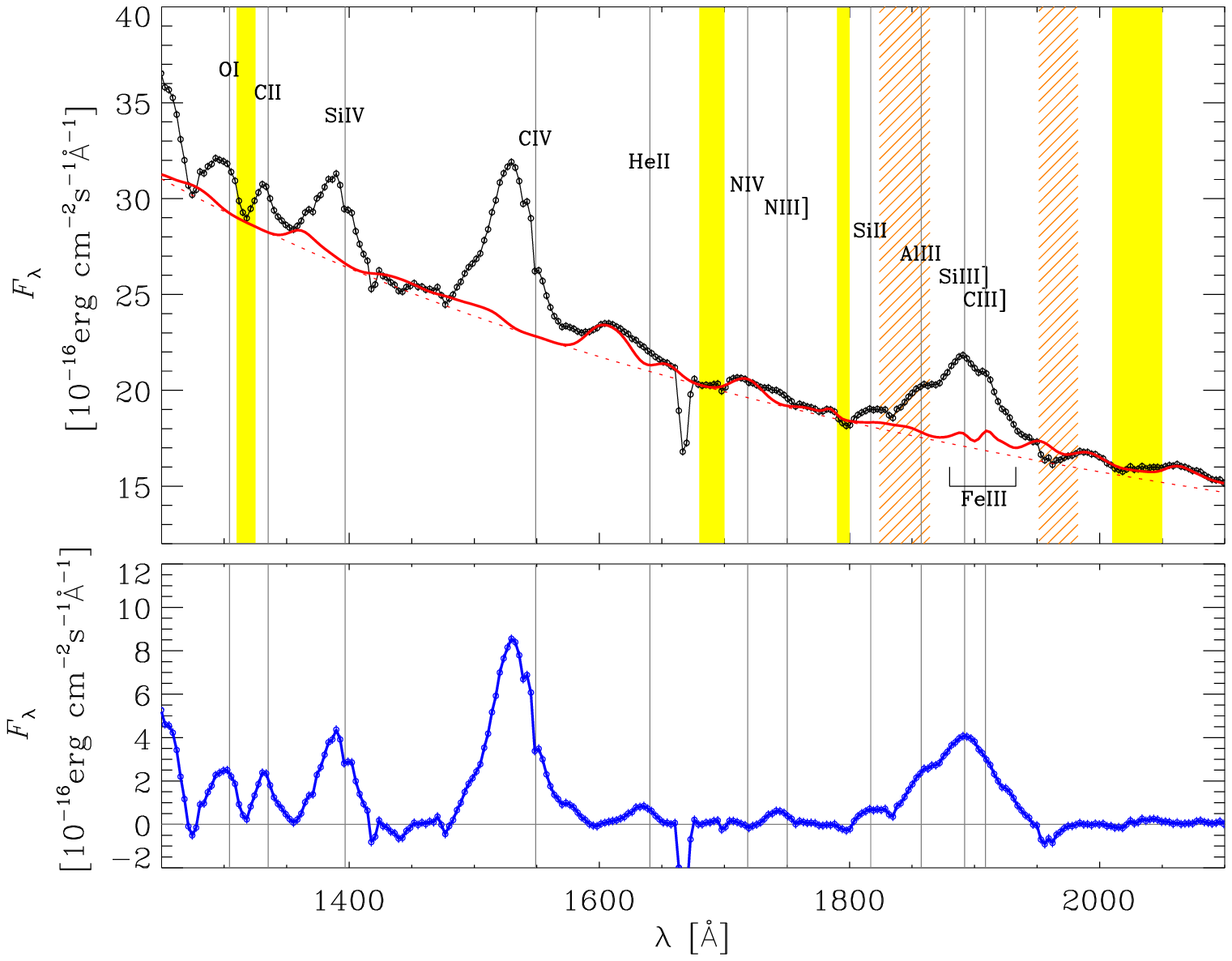}
    \caption{The central part of the mean optical dereddened spectrum, shifted to the rest frame according to $z=2.213$. Top: the black points show the spectrum before the continuum and Fe subtraction. The continuum (red dotted line) is obtained by fitting the spectral zones marked with yellow stripes with a power law.
The zones in yellow were also used to estimate the mean flux uncertainty. 
The red solid line highlights the Fe contribution, which is obtained by scaling and broadening the \citet{vestergaard2001} Fe template. 
The contribution of blueshifted \ion{Fe}{III} is indicated.
Slanted orange stripes 
indicate atmospheric absorption.
Bottom: the blue points show the clean spectrum, after removal of Fe and continuum. Vertical grey lines mark the position of the main AGN emission lines at $\lambda_{\rm lab}$. }
    \label{fitto}
\end{figure}

The central part of the optical mean spectrum corrected for Galactic reddening and shifted to the rest frame is plotted in Fig.\ \ref{fitto}. The positions of the telluric absorption lines are indicated, together with those of the main emission lines found in AGN.
Four small zones were selected to estimate the continuum and the mean flux uncertainty. 
The slope of the continuum is $\alpha_\lambda=-1.42$ ($\alpha_\nu=-0.58$), which is in the range of the values found by other authors \citep[see][for a review]{vandenberk2001}.

Modelling the Fe contribution is not an easy task, especially in the \ion{C}{III]} line region.
A small bump in the red part of \ion{C}{III]} was ascribed to blueshifted \ion{Fe}{III}.
After continuum and Fe subtraction, the mean spectrum shows, besides the major broad emission lines \ion{C}{III]}, \ion{C}{IV}, \ion{Si}{IV}, \ion{C}{II} and \ion{O}{I}, also two minor bumps likely corresponding to \ion{He}{II} $\lambda$1640 and  \ion{N}{III]} $\lambda$1750. Both the major and these minor lines are clearly blueshifted with respect to the position they would have according to the systemic redshift. 

We also note that the high-ionization lines \ion{Si}{IV} and \ion{C}{IV} 
show a little dip in the profile close to the systemic redshift. This feature is present in all single spectra, both from the NOT and WHT. While the feature is tiny, its ubiquity suggests that it corresponds to a source property. We speculate that it may be due to self-absorption.

\subsection{C III}
\begin{figure}
	\includegraphics[width=\columnwidth]{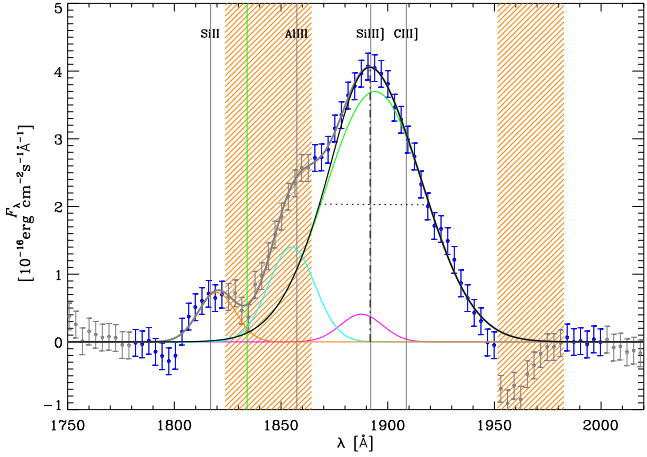}
    \caption{Spectral fitting of the mean \ion{C}{III]} line shown in the bottom panel of Fig.\ \ref{fitto}. The grey vertical lines indicate the position of the features at $\lambda_{\rm lab}$.
Because of absorption by telluric (orange slanted stripes) 
and Galactic (green) lines and blending by \ion{Al}{III}, part of the profile marked by grey dots cannot be used for spectral fitting. Only the blue points are considered in the model fit. This includes four Gaussians, two of which to reproduce the \ion{C}{III]} line (green and magenta). The total model fit to the spectrum is represented by the solid thick grey line, while the black one highlights the \ion{C}{III]} line profile.
The dotted-dashed black line marks the line peak $\lambda^{\rm peak}_{\rm rest}$ according to the model; the dotted horizontal line highlights the FWHM.}
    \label{fitto_ciii}
\end{figure}

Modelling the \ion{C}{III]} line is a difficult task because of blending and absorption.
The line peak in the rest frame roughly coincides with the laboratory wavelength of \ion{Si}{III}]. However, in average quasar spectra the flux of \ion{Si}{III}] is about one hundredth that of \ion{C}{III}], just as \ion{Al}{III}, on the blue side of the line, represents only $\sim 2\%$ of the \ion{C}{III} line flux \citep{vestergaard2001}.
Therefore, we are confident that the line peak, as well as most of the line flux, actually refers to blueshifted \ion{C}{III}].
We masked the spectral region corresponding to telluric absorption and fitted the remaining profile with four Gaussians (see Fig.\ \ref{fitto_ciii}). Two of them are used to reconstruct the blue part, where blending with \ion{Al}{III} is present. The result must be considered as a reasonable guess. 
The other two Gaussians define the \ion{C}{III]} profile.

\subsection{C IV}
\begin{figure}
	\includegraphics[width=\columnwidth]{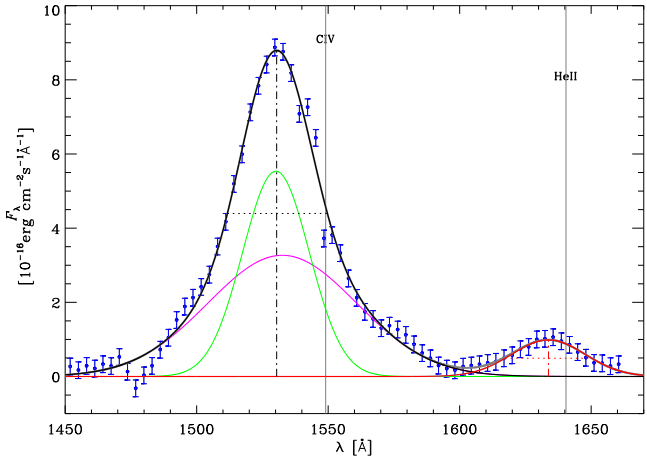}
    \caption{Spectral fitting of the mean \ion{C}{IV} line shown in the bottom panel of Fig.\ \ref{fitto}. Two Gaussians (green and magenta) are used to reproduce the \ion{C}{IV} profile (black); a third Gaussian (red) accounts for the little bump on the red side due to \ion{He}{II}.
The grey vertical lines indicate the positions of \ion{C}{IV} and \ion{He}{II} at $\lambda_{\rm lab}$.
The dotted-dashed lines mark the line peak $\lambda^{\rm peak}_{\rm rest}$ according to the models, the dotted horizontal lines highlight the FWHM.}
    \label{fitto_civ}
\end{figure}

The \ion{C}{IV} line is particularly useful in spectroscopic studies because it is strong and not affected by significant blending.
Spectral fitting to the \ion{C}{IV} line is shown in Fig.\ \ref{fitto_civ}. The final model includes three Gaussians: two for the \ion{C}{IV} line and one to reproduce the little bump ascribed to \ion{He}{II} emission. 

\subsection{Si IV, C II, O I}
\begin{figure}
	\includegraphics[width=\columnwidth]{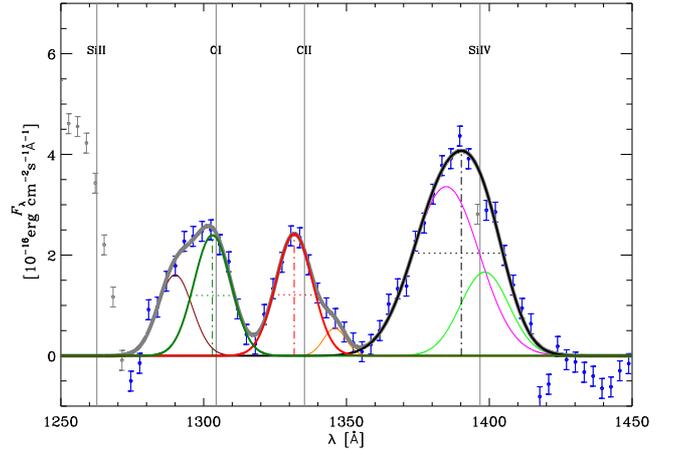}
    \caption{Spectral fitting of the mean \ion{Si}{IV}, \ion{C}{II}, and \ion{O}{I} lines shown in the bottom panel of Fig.\ \ref{fitto}. The grey vertical lines indicate the position of the features at $\lambda_{\rm lab}$.
Grey points represent the whole mean spectrum in the region, blue points indicate the data points used for the spectral fitting.
The total model (solid thick grey line) includes six Gaussians.
The models for the main emission lines are shown in black, red and dark green for \ion{Si}{IV}, \ion{C}{II}, and \ion{O}{I}, respectively. The \ion{Si}{IV} line model involves two Gaussians (green and magenta).
The dotted-dashed lines mark the peaks $\lambda^{\rm peak}_{\rm rest}$ according to the models; the dotted horizontal lines indicate the FWHM.}
    \label{fitto_siiv}
\end{figure}

When we speak of \ion{Si}{IV} we actually mean  \ion{Si}{IV}+\ion{O}{IV]}.
The \ion{Si}{IV}, \ion{C}{II}, and \ion{O}{I} lines must be treated together, as they are very close. The blue wings of \ion{Si}{IV} and \ion{O}{I} are blended by Fe. The continuum- and Fe-subtracted spectrum is shown in Fig.\ \ref{fitto_siiv}. We model this spectral region with six Gaussians. Two of them account for the \ion{Si}{IV} profile, whose asymmetry is likely due to blending with \ion{O}{IV]}. Two Gaussians reproduce the \ion{C}{II} and \ion{O}{I} lines. The remaining two Gaussians account for the flux excess blueward and redward of the \ion{O}{I} and \ion{C}{II} lines, respectively, which are possibly due to non-iron weak features \citep[see][]{vestergaard2001}. 

\section{Line flux ratios}
\label{ratios}

\begin{figure}
	\includegraphics[width=\columnwidth]{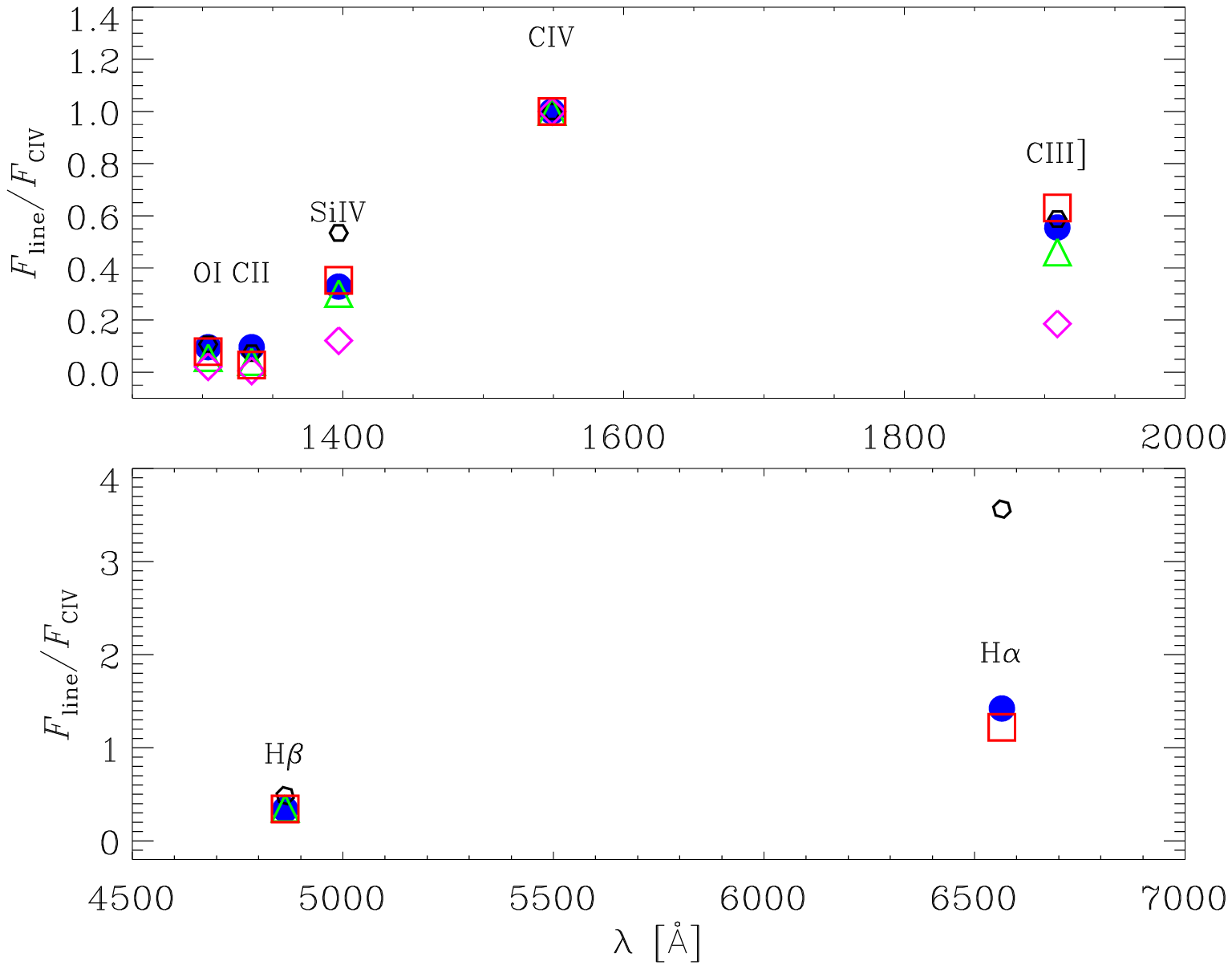}
    \caption{Line flux ratios with respect to \ion{C}{IV} in 4C~71.07 (blue dots) and in various composite quasar spectra: \citet[][black hexagons]{cristiani1990}, \citet[][green triangles]{francis1991}, \citet[][magenta diamonds]{zheng1997}, \citet[][red squares]{vandenberk2001}.}
    \label{fratio_l}
\end{figure}

AGN show typical flux ratios between lines, even if with large spreads around the mean.
In order to see whether the properties of 4C~71.07 are compatible with those of QSO, in Fig.\ \ref{fratio_l} we compare our results with the mean flux ratios of various composite quasar spectra \citep{cristiani1990,francis1991,zheng1997,vandenberk2001}.

We chose \ion{C}{IV} as normalisation because it was present in all datasets. Our values lie in the range or close to the literature ones. 

We note in particular the strength of \ion{H}{$\alpha$} in the dataset of \citet{cristiani1990}, consisting of a small number of bright objects.

\section{Black hole mass}
\label{bh}

The best technique to estimate the mass of the central black hole in AGN is reverberation mapping \citep[e.g.][]{kaspi2000,peterson2004,bentz2009b}. By monitoring the trends of both lines and their photoionizing continuum, and by running cross-correlation analysis, one can estimate the time delay with which lines react to variations of the continuum and, in turn, the size of the BLR. From the BLR size and gas velocity derived from line broadening, it is then possible to estimate the black hole mass by means of the virial equation $M_{\rm BH}=R_{\rm BLR} \, v_{\rm BLR}^2/G$, where $R_{\rm BLR}=c \, \tau$ is the BLR radius, $\tau$ the time delay of the line flux variations after the continuum flux changes in the rest frame, and $v_{\rm BLR}$ is the gas velocity.

However, reverberation mapping requires spectral monitoring for long periods of time. Therefore, scaling relationships have been derived that allow us to estimate the black hole mass even with single-epoch spectra \citep[e.g.][]{kaspi2005,bentz2009a}.
They rely on Balmer lines for nearby objects, while for high-redshift sources rest-frame ultraviolet lines are used, namely \ion{Mg}{II} and \ion{C}{IV} \citep{vestergaard2006,mclure2002}.

In general, these methods cannot be applied to derive black hole masses for blazars, where the continuum is usually dominated by the synchrotron non-thermal emission from the jet, which is not the source of the ionizing photons, or at least not the main one (see Section \ref{lineva}). In the case of 4C~71.07, the optical emission is dominated by the big blue bump due to the disc thermal emission \citep{raiteri2019}, so in principle the application of methods developed for other AGN can grossly work. Moreover, the model developed by \citet{raiteri2019} allows us to separate the thermal from the non-thermal emission contributions.

The orientation of blazars is known: their jet is pointing towards us and the accretion disc is viewed face-on. As a consequence the BLR, which is likely to have a flattened structure \citep{mclure2004, decarli2011}, is seen face-on too. 
A big issue in determining the black hole mass of AGN in general, and of 4C~71.07 in particular, is the presence of outflows and thus the separation of the outflow component from the rotational one.
The problem has been tackled by a number of authors. As we saw also in the case of 4C~71.07, the \ion{C}{IV} broad emission line is particularly affected by blueshift due to outflows \citep{sulentic2007,ge2019} and cannot be used as a virial broadening estimator, unless some correction is applied \citep{runnoe2013,coatman2017,marziani2019}.

In the following we apply different methods and compare different black hole mass estimates.
We consider the mass scaling relationships by \cite{vestergaard2006}, \cite{coatman2017}, and \citet{shen2011}, which use the \ion{H}{$\beta$}, \ion{C}{IV} or \ion{H}{$\alpha$} emission lines to determine the black hole mass through their FWHM and either the continuum or line luminosity:

$$\log M_{\rm BH}= a \, \log b + c \, \log d + e$$

where $a$, $c$ and $e$ are constants, $b=\rm FWHM/(1000 \, km \, s^{-1})$, and $d$ is either the continuum\footnote{Estimated at 5100 \AA\ for \ion{H}{$\beta$} and at 1350 \AA\ for \ion{C}{IV}.}  or line luminosity, i.e.\ either$(\lambda \, L_\lambda)/(10^{44} \rm \, erg \, s^{-1})$ or 
$L_{\rm line}/(10^{42} \rm \, erg \, s^{-1})$.
In the case of \ion{C}{IV}, the prescriptions by \citet{coatman2017} include a correction of the FWHM for the outflow contribution.

To estimate the photoionizing continuum luminosity in the optical and near-infrared bands, we adopted the model by \citet{raiteri2019}, who separated the emission contribution of the disc from that of the jet. From their work we can infer that while the flux at 1350 \AA\ is dominated by thermal emission, which amounts to $\sim 80\%$ of the total flux, at 5100 \AA\ the two components contribute roughly the same quantity. 
Table \ref{mass} reports the input values we used and the corresponding black hole masses derived. 
Those estimated from the \ion{H}{$\beta$} and \ion{H}{$\alpha$} lines and from the \ion{C}{IV} line after blueshift correction are within a factor 2, which we consider a fair agreement. These results would indicate a black hole mass of $M_{\rm BH}=(2.0 \pm 0.7) \times 10^9 \, M_\odot$ for 4C~71.07. 
We note that the correction for blueshift applied to \ion{C}{IV} implies a virial FWHM only $\sim 10\%$ larger than the FWHM of \ion{H}{$\beta$}.  

In \citet{raiteri2019} the empirical model for the thermal emission in 4C~71.07 allowed us to calculate a disc bolometric luminosity $L_{\rm disc}=2.45 \times 10^{47} \rm \, erg \, s^{-1}$. By assuming a \citet{shakura1973} accretion disc, we used the \citet{calderone2013} laws to derive the mass accretion rate, $\dot{M} \simeq 18 \, M_\odot  \rm \, yr^{-1}$, and black hole mass, $M_{\rm BH} \simeq 1.6 \times 10^9 \, M_\odot$, from it. 
This value of the black hole mass is consistent with those obtained above from the \ion{H}{$\beta$},  \ion{H}{$\alpha$}, and \ion{C}{IV} lines.
Our result is also compatible with \citet{ghisellini2010}, who inferred a black hole mass of $3 \times 10^9 \, M_\odot$ using a simple leptonic, one-zone synchrotron and inverse Compton model to fit the SED.

In contrast and as expected, the black hole mass obtained from the \ion{C}{IV} line without blueshift correction is almost five times higher and confirms the danger of using standard scaling relationships when outflows are present.

From the black hole mass we can derive the Eddington luminosity $L_{\rm Edd} = 1.26 \times 10^{38} M_{\rm BH}/M_\odot \rm \, erg \, s^{-1}  = 2.49 \times 10^{47} \rm \, erg \, s^{-1}$, and the Eddington ratio $L_{\rm disc}/L_{\rm Edd} \approx 1$. This means that the force of radiation pressure generated by accretion on the ionized matter has approximately the same strength of the gravitational force due to the supermassive black hole. Only a few quasars show such a high Eddington ratio \citep{sun2018}.

\begin{table*}
	\caption{Black hole mass estimates according to various scaling relations.}
	\label{mass}
	\begin{tabular}{cccccccc}
\hline
\multicolumn{8}{c}{Line luminosity}\\
Line              & FWHM & $L$  & a      & c    & e     & ref  & $M_{\rm BH}$ \\
                  & $\rm km \, s^{-1}$ & $\rm 10^{44} \, erg \, s^{-1}$  &      &   &  & & $10^9 \, M_\odot$ \\
\hline
\ion{H}{$\beta$}  & 3240 &  5.054  & 2      & 0.63 & 6.67  & VP06 & 2.5 \\
\ion{H}{$\alpha$} & 3131 &  21.31  & 2.1    & 0.43 & 6.679 & S11  & 1.4 \\
\hline
\multicolumn{8}{c}{Continuum luminosity}\\
Line     & FWHM & $\lambda \, L_\lambda$ & a & c & e & ref & $M_{\rm BH}$ \\
         & $\rm km \, s^{-1}$ & $\rm 10^{46} \, erg \, s^{-1}$    &   &   &  &  & $10^9 \, M_\odot$ \\
\hline
\ion{H}{$\beta$} & 3240 &  2.726 & 2 & 0.50 & 6.91 & VP06 & 1.4 \\
\ion{C}{IV}      & 7451 &  11.15 & 2 & 0.53 & 6.66 & VP06 & 10.4 \\
\ion{C}{IV}      & 3549 &  11.15 & 2 & 0.53 & 6.71 & C17  & 2.6 \\
\hline
\end{tabular}

VP06=\citet{vestergaard2006};\\
S11=\citet{shen2011};\\
C17=\citet{coatman2017}\\
\end{table*}

\section{BLR properties}
\label{blr}

The FWHM velocities reported in Table \ref{results} suggest a stratified BLR, where different lines are emitted at different distances from the supermassive black hole. The high-ionization species, with larger $v_{\rm FWHM}$, are likely produced in inner zones than the low-ionization lines. Indeed, the stratification of the BLR is also confirmed by reverberation mapping (see also Sect.\ \ref{bh}), as different lines lead to different values of the BLR radius.

From the relation between the radius of the BLR, $R_{\rm BLR}$, and the luminosity of the continuum at 5100 \AA\ given by \citet{mclure2004}, we can estimate that $R_{\rm BLR} \sim 800$ light days at the distance where the \ion{H}{$\beta$} is produced. 
If we use the relation between $R_{\rm BLR}$ and the continuum luminosity at 1350 \AA\ given by \citet{grier2019}, we obtain $R_{\rm BLR} \sim 700$  light days for the \ion{C}{IV} line. 
The large scatter affecting the above relations imply uncertainties of the order of a few hundred light days on $R_{\rm BLR}$. 
Therefore, we cannot verify the hypothesis of a stratified BLR.

The BLR luminosity can be estimated as the total luminosity of the broad emission lines.
Adopting the relative fluxes reported by \citet{vandenberk2001}, we calculate the total flux of the broad emission-line features
and the ratios of the main emission lines fluxes with respect to the total flux.
We use these ratios and the isotropic luminosities of the main lines listed in Table \ref{results} to calculate the BLR luminosity of 4C~71.07. We obtain the values (1.47, 1.51, 1.40, 1.49, 1.76) $\times 10^{46} \rm \, erg \, s^{-1}$ from \ion{Si}{IV}, \ion{C}{IV}, \ion{C}{III]}, \ion{H}{$\beta$} and \ion{H}{$\alpha$}, respectively, whose mean and standard deviation give $L_{\rm BLR}=(1.52 \pm 0.14) \times 10^{46} \rm \, erg \, s^{-1}$.
This is a very high value when compared to those of other blazars and AGN in general \citep[e.g.][]{ghisellini2011, sbarrato2014}.
By considering that $L_{\rm disc} \approx L_{\rm Edd} \approx 2.5 \times 10^{47} \rm \, erg \, s^{-1}$ (see Sect.\ \ref{bh}), we derive that $L_{\rm BLR}$ is about 6\% of both the disc and the Eddington luminosity. This result places 4C~71.07 in the upper tail of the distribution of $L_{\rm BLR}/L_{\rm Edd}$ values found by \citet{sbarrato2014} by analysing a large sample of blazars and radio galaxies.

\section{Blueshifts and outflow properties}
\label{blue}

\begin{figure}
	\includegraphics[width=\columnwidth]{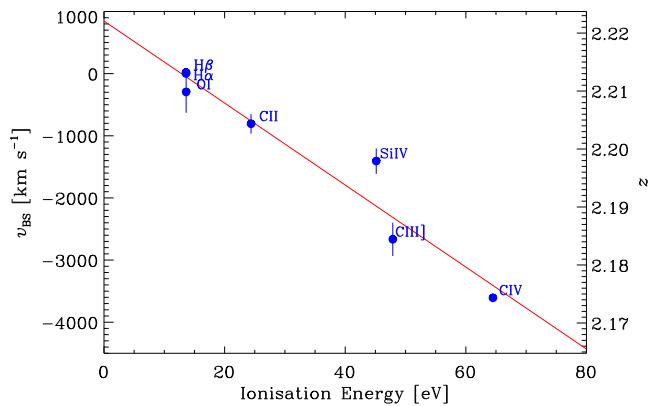}
    \caption{Blueshift velocities of the main broad emission lines in 4C~71.07 versus the ionization energy of the corresponding species.
The vertical scale on the right gives the corresponding redshift. 
The red line represents the least-squares linear regression line. }
    \label{bs_ione}
\end{figure}

It is known that a correlation exists between the line shift and the ionization potential \citep[e.g.][]{vandenberk2001}. Figure \ref{bs_ione} shows the blueshift velocities reported in Table \ref{results} as a function of the ionization energy. 
On the right scale we see how the source redshift decreases when it is measured from higher excitation lines. 
The largest blueshift velocity is shown by the \ion{C}{IV} line and exceeds $\rm 3600 \, km \, s^{-1}$, a very high value if compared to those characterising type 1 AGN \citep{vandenberk2001,vietri2018}. Large blueshifts of the \ion{C}{IV} line are found in Population A AGN sources, defined as having \ion{H}{$\beta$} FWHM less than $\rm 4000 \, km \, s^{-1}$ \citep{sulentic2007}, and in sources with large Eddington ratios \citep{sun2018}. 4C~71.07 shares both characteristics.

\begin{figure}
	\includegraphics[width=\columnwidth]{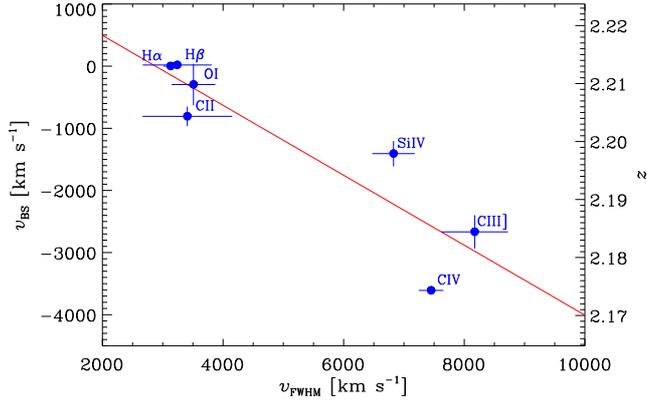}
    \caption{Line blueshift versus broadening velocity of the main emission lines in 4C~71.07.
The vertical scale on the right gives the corresponding redshift.
The red line represents the least-squares linear regression line.}
    \label{bs_v}
\end{figure}

4C~71.07 is a blazar, which means that the accretion disc is viewed nearly face-on. If the BLR had a flattened geometry, we would expect to observe only a small enlargement of the broad emission lines, as only a small component of the rotational motion of the gas clouds around the supermassive black hole would be towards the line of sight. 
The high-excitation broad emission lines of 4C~71.07 are instead rather wide, which implies that either the part of BLR producing them does not have a flattened geometry or the velocity we measure is mainly due to outflows. 
The strong blueshifts we derived imply that this is actually the case. Consequently, we would expect a correlation between the line blueshift and broadening. This correlation is evident in Fig.\ \ref{bs_v}. 
A similar effect was found in quasars at both low and high redshifts \citep[e.g.][]{negrete2018,vietri2018}. However, as far as we know, this is the first time that the correlation is found for a blazar.

\begin{figure}
	\includegraphics[width=\columnwidth]{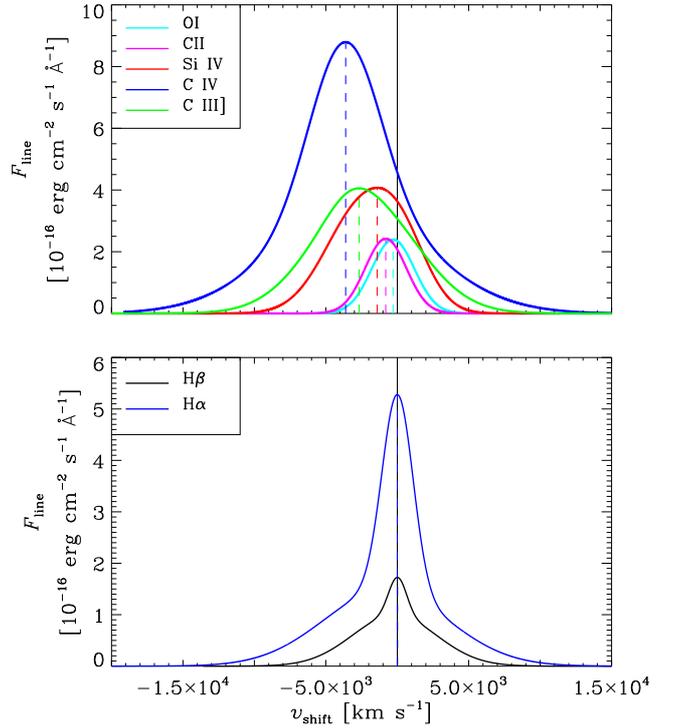}
    \caption{Line velocity profiles of the main broad emission lines in the 4C~71.07 spectra with respect to the systemic redshift $z=2.213$ estimated from the Balmer lines. The peaks of the high-excitation lines occur at $v_{\rm shift}=v_{\rm BS}$. In the top panel, from left to right we find \ion{C}{IV}, \ion{C}{[III]}, \ion{Si}{IV}, \ion{C}{II}, and \ion{O}{I}. In the bottom panel, \ion{H}{$\alpha$} (upper line) and \ion{H}{$\beta$} (lower line).
 }
    \label{vshift}
\end{figure}

Figure \ref{vshift} shows the line profiles as a function of the velocity shift, defined as:

\begin{equation}
v_{\rm shift}(\lambda)=c \, {{\lambda - \lambda_{\rm sys}} \over {\lambda_{\rm sys}}} 
\end{equation}
For $\lambda=\lambda^{\rm peak}_{\rm obs}$ we get $v_{\rm shift}=v_{\rm BS}$. 

The plot uses the spectral fits obtained in the previous sections to the Fe- and continuum-subtracted mean de-reddened spectra. 
The shape of the Balmer lines shows a narrow core and wide wings. As already noticed, \ion{Si}{IV} presents a slightly asymmetric profile, while the other lines are more symmetric.
Differences in line profiles in the same object is not a rare phenomenon.

In a recent paper, \citet{fiore2017} presented AGN wind scaling relations.
If we consider the 4C~71.07 bolometric luminosity estimated by \citet[][see Sect.\ \ref{blr}]{raiteri2019}, these scaling relations for ionized winds in high-redshift objects would lead to extremely high mass outflow rates, up to several thousand solar masses per year and to outflow kinetic powers up to $10^{46} \, \rm erg \, s^{-1}$. The maximum wind velocities in 4C~71.07 appear at the highest limit of the measured range for ionized winds.

\section{Line variability}
\label{lineva}

As already mentioned, in unbeamed AGN the broad emission lines are assumed to be produced in a BLR photoionized by radiation from the accretion disc, which we measure as continuum. 
In blazars the optical and even more the near-infrared continua are usually dominated by synchrotron radiation from the jet.
In principle one might expect that the jet can contribute to the photoionization of the BLR
if the jet dissipation zone is embedded in the BLR \citep{leon2013}. If this is the case, we should see an increase of the line flux when the source is in a flaring state, unless the activity is due to geometrical reasons, i.e.\ to a better alignment of the emission region with the line of sight  \citep[e.g.][]{raiteri2017_nature}.

If we look at the past literature, we can find a number of blazar spectroscopic monitoring works that in most cases concluded that the BLR does not seem to react to jet activity \citep[e.g][]{corbett2000,raiteri2007a,carnerero2015,isler2015}.
However, in a few cases some increase of the broad emission lines flux has been detected \citep{leon2013,isler2013,isler2015}. 
All the above results would suggest that usually the jet does not contribute to the BLR ionization, likely because the dissipation zone is outside the BLR. However, if for some reason the beamed radiation from the jet illuminates part of the BLR, 
then we should see the correlation between the jet and line fluxes.

The optical and near-infrared emission of 4C~71.07 is dominated by thermal radiation from the disc \citep{raiteri2019}. During the period of our spectroscopic monitoring, some flaring episodes have been revealed, which are expected to come from the jet activity because of the short variability time scales and correlated $\gamma$-ray and X-ray flux changes \citep{vercellone2019,raiteri2019}. Therefore, we can investigate whether the broad line fluxes show some correlation with the jet emission. 

Because of the smaller S/N with respect to the average spectrum analysed before, we fit the single-spectra lines with single Gaussians, without Fe subtraction or line deblending. In this way we investigate whether the BLR in general reacts to changes of the continuum flux. Here the continuum is estimated locally and the line flux depends on the local shape of the spectrum. Therefore, we calculate uncertainties by adding in quadrature the errors obtained with the usual Monte Carlo flux redistribution method and the errors obtained by performing both a linear and quadratic fit to the continuum. We consider only the strongest lines, \ion{C}{IV} and \ion{C}{III]}.

\begin{figure}
	\includegraphics[width=\columnwidth]{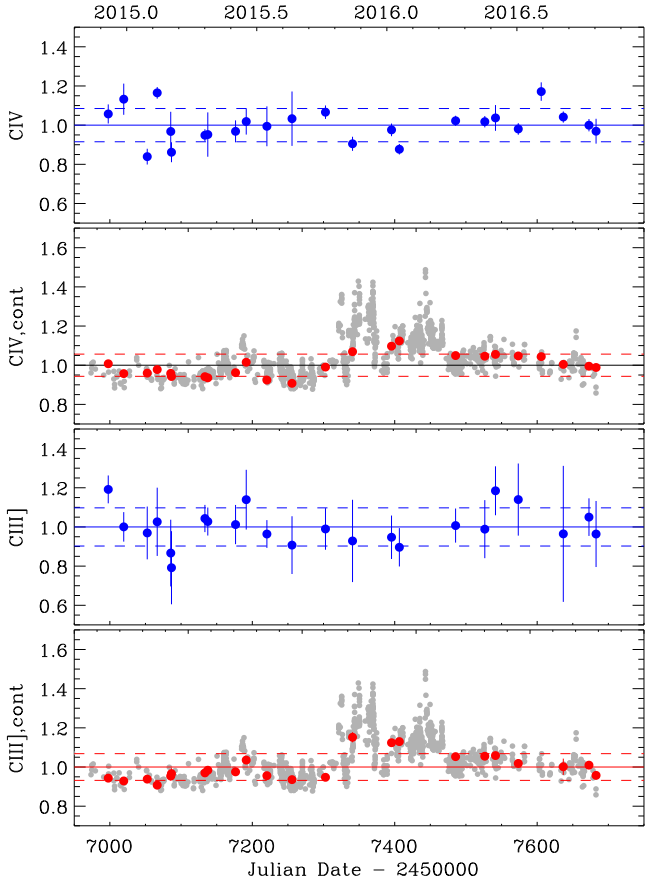}
    \caption{Behaviour of line and continuum fluxes as a function of time for \ion{C}{IV} and \ion{C}{III]}. Fluxes are normalized to the mean value.
The grey dots in the continuum plots show the best-sampled $R$-band light curve obtained by the WEBT Collaboration and presented by \citet{raiteri2019}. It has been normalized to the mean value and then scaled by $+0.05$ to match the spectra continuum.}
    \label{flussi_righe}
\end{figure}

The behaviour of the line and continuum fluxes as a function of time is displayed in Fig.\ \ref{flussi_righe}. 
The trends of the continuum flux corresponding to the \ion{C}{IV} and \ion{C}{III]} lines are very similar and fairly match the behaviour of the well sampled $R$-band light curve obtained by the WEBT Collaboration \citep{raiteri2019}. In particular, the flux increase between $\rm JD=2457300$ and 2457400 marks the jet flaring activity analysed by \citet{vercellone2019} and \citet{raiteri2019}. Neither the \ion{C}{IV} nor the \ion{C}{III]} line fluxes show any clear response to this activity. 
Actually, in the three spectra acquired during the continuum flaring period (JD $\sim$ 2457340--7410), the line fluxes are below their average values. 
Therefore, it seems that the jet does not contribute to the BLR ionization, at least in the time span considered here. 

Any reaction of the line fluxes to the thermal continuum changes cannot be verified, as it requires much longer time-scales. Indeed, in Sect.\ \ref{blr} we estimated BLR radii of about 2 light years, which implies that the expected time delay of the line flux variations after the thermal flux changes in the observed frame is of the order of 6 years.

\section{Discussion and conclusions}
\label{fine}
In this paper we have investigated the unbeamed nuclear properties of the FSRQ 4C~71.07 through a detailed analysis of its optical and near-infrared spectra. 
This blazar is known for the extreme properties of its beamed radiation from the jet, including high radio, X-ray and $\gamma$-ray luminosity, strong flux and polarization variability, and high Compton dominance \citep[e.g.][]{raiteri2019}.
Our goal here was to unveil the nature of its quasar core.

As expected from high-redshift sources viewed face-on, we have found no evidence for NLR contribution in the spectra.
Therefore, we have estimated the systemic redshift from the Balmer lines \ion{H}{$\beta$} and \ion{H}{$\alpha$}, obtaining $z_{\rm sys}=2.2130 \pm 0.0004$. 
 
All the other primary broad emission lines appear blueshifted, which is an indication of outflow. The estimated outflow velocities can reach more than 3000 $\rm km \, s^{-1}$. These are rather high values when compared to those of other AGN \citep[e.g.][]{vandenberk2001, vietri2018}.
We have found that the amount of both line blueshift and broadening depends on the ionizing energy, with high-ionization lines showing larger blueshifts and velocities. This means that the outflow is also responsible for the line broadening of at least the high-ionization species.

The picture is that of a stratified BLR, with radii of several hundred light days, and which is strongly affected by ionized winds. The velocity of these winds decreases going from the inner regions, where the high-ionization lines are produced, to the outer regions, from which the low-ionization lines come from.

We have applied different scaling relations to estimate the mass of the central black hole, based on the \ion{H}{$\beta$}, \ion{H}{$\alpha$}, and \ion{C}{IV} lines.  
In the last case we also corrected the line FWHM for the effect of  blueshift according to the prescriptions of \citet{coatman2017}. This leads to a virialized FWHM of \ion{C}{IV} close to that of the Balmer lines.
Although the presence of outflows makes the use of methods based on the virialization assumption somewhat hypothetical, we have obtained a value, $M_{\rm BH} = (2.0 \pm 0.7) \times 10^9 \, M_\odot$, which is consistent with the estimates inferred from other methods.
The fact that different methods to calculate the black hole mass yield comparable results suggests that the FWHM of the lines do not need a large correction for orientation.
This may be surprising, because  blazars usually show narrower broad lines than other type 1 AGN, as expected for flattened BLRs viewed face-on \citep{decarli2011}.
One explanation may be that the BLR is not particularly flat.
Alternatively, the orientation correction, which would increase the FWHM of the lines, may be compensated for by an FWHM narrowing correction, possibly due to outflow even affecting the Balmer lines.
Although we cannot distinguish between the two alternatives, the coincidence of a double correction of nearly the same amount makes us favour a scenario where the velocity we measure from the Balmer lines is not significantly affected by outflow.

From the black hole mass we have derived an Eddington luminosity $L_{\rm Edd} = 2.49 \times 10^{47} \rm \, erg \, s^{-1}$ and an Eddington ratio $L_{\rm disc}/L_{\rm Edd} \approx 1$, which means that the radiation and gravitational forces are of the same order. 
We have estimated a BLR luminosity $L_{\rm BLR} = (1.52 \pm 0.14) \times 10^{46} \rm \, erg \, s^{-1}$, which represents about 6\% of the disc and Eddington luminosities. This identifies the quasar core of 4C~71.07 as one of the most luminous among the blazar nuclei \citep{ghisellini2011,sbarrato2014}. 

Finally, we have investigated line flux variability. In the considered period (2014 December -- 2016 October) there is no significant line variability, not even during the jet flaring activity detected in the optical, X-ray and $\gamma$-ray bands in 2015 October--November. Therefore, we conclude that at that time the dissipation zone in the jet was likely outside the BLR. Alternatively, if the flaring activity is not due to jet intrinsic processes, but to a better alignment of the jet emitting zone with the line of sight, the line fluxes would not be affected, whatever the position of the BLR relative to the jet. The detection of thermal continuum variability in this high-redshift object likely requires much longer time-scales than that considered here. Therefore, we could not study the expected delayed reaction of the emission lines to continuum flux changes, which would take about 6 years.

In conclusion, we have detected extreme outflow properties in a blazar with extreme nuclear and jet emission properties. In a forthcoming paper we will extend this study to a sample of high-redshift blazars to see if and how these components are linked.

\section*{Acknowledgements}
We thank Marianne Vestergaard for sending us the UV iron templates published in \citet{vestergaard2001}, and Alessandro Capetti for useful discussions.
We acknowledge financial contribution from the agreement ASI-INAF n.2017-14-H.0 and from the contract PRIN-SKA-CTA-INAF 2016.
Partly based on observations made with the Italian Telescopio Nazionale Galileo (TNG) operated on the island of La Palma by the Fundaci\'on Galileo Galilei of the INAF (Istituto Nazionale di Astrofisica) at the Spanish Observatorio del Roque de los Muchachos of the Instituto de Astrofisica de Canarias.
Partly based on observations made with the Nordic Optical Telescope, operated by the Nordic Optical Telescope Scientific Association at the Observatorio del Roque de los Muchachos, La Palma, Spain, of the Instituto de Astrofisica de Canarias.
The NOT data presented here were obtained with ALFOSC, which is provided by the Instituto de Astrofisica de Andalucia (IAA) under a joint agreement with the University of Copenhagen and NOTSA. ALFOSC observations were done as part of the service observation programs P51-008 and P53-017. 
Partly based on data acquired within the service programs SW2014b14 and SW2015b13 using the instrument ACAM at the William Herschel telescope (WHT). The WHT and its service programme are operated on the island of La Palma by the Isaac Newton Group of Telescopes in the Spanish Observatorio del Roque de los Muchachos of the Instituto de Astrof\'isica de Canarias.






\bsp	
\label{lastpage}
\end{document}